# Real time magneto-optical imaging of vortices in superconductors

Pål Erik Goa*, Harald Hauglin*, Michael Baziljevich*, Eugene Il'Yashenko*, Peter L. Gammel†, Tom H. Johansen*

*Department of Physics, University of Oslo, PO Box 1048 Blindern, 0316 Oslo, Norway.
†Bell Laboratories, Lucent Technologies, Murray Hill, New Jersey 07974, USA

The dynamic behaviour of the quantized magnetic vortices in type-II superconductors is of great interest. It determines the critical current density of superconductors[1] and can also serve as a model for condensed matter flow[2]. We demonstrate here real-time imaging of individual vortices in a $NbSe_2$ single crystal using polarized light microscopy. A new high-sensitivity magneto-optical (MO) imaging system enables observation of the static vortex lattice as well as single vortex motion at low flux densities.

Several methods for individual vortex visualization already exist[3]. Examples are the techniques of Bitter decoration[4], of scanning magnetic probes[5-7] and of Lorentz microscopy[8]. The main advantage of our method is the high temporal resolution combined with its applicability to any superconducting sample with a flat surface. This, together with the relative simplicity of an MO imaging setup, mean that the method complements existing techniques and opens new possibilities for vortex dynamics studies.

As in conventional MO imaging[9], we employ plane polarized light and the Faraday effect in a ferrite garnet film (FGF) to visualize the local magnetic field over the superconductor surface (Fig. 1a). The main challenges for individual vortex observation are to resolve the magnetic field modulation decaying rapidly with distance from the sample surface, and to minimize depolarization effects in the optical system leading to loss of polarization contrast. To this end we have constructed a combined cryostat/MO system with a modular open microscope featuring a 100 W Hg lamp, an Olympus LMPlan 50x objective mounted inside a modified Hi-Res (Oxford Instruments) He flow cryostat, a Glan-Taylor polarizer/analyzer pair, a Smith beam splitter and a cooled CCD camera.

A cleaved $NbSe_2$ single crystal[10] (3x2x0.1 $mm^3$) with $T_C$ = 7.2 K was mounted in the cryostat using vacuum grease, and a 0.8 μm thick FGF grown by liquid phase epitaxy on a Gadolinium Gallium Garnet (GGG) substrate was placed on top of the superconductor by applying a small mounting pressure. The FGF, with the chemical composition $(Bi,Lu)_3(Fe,Ga)_5O_{12}$, has a small-field Faraday coefficient of 8.3 °/μm·kOe for light at 546 nm wavelength. In order to minimize the distance between the sample surface and the MO layer, we omit a separate reflective layer and use the sample itself as a mirror.

Typical results are shown in Fig. 1. Images of the vortex lattice (panel **b** and **c**) are obtained by subtracting two raw images recorded with the analyzer rotated 88 and 92 degrees relative to the polarizer. This differential scheme increases the signal-to-noise ratio and partly compensates for variations in the reflectivity of the sample. At present we can resolve the individual vortices up to a maximum flux density of 1.0 mT, corresponding to an intervortex distance of 1.4 µm.

Vortex dynamics can be studied by recording a time series of MO images. The temporal resolution is currently limited to 10 frames/sec by the image acquisition system. Details of vortex motion can be visualized by subtracting images taken at different points in time. The contrast in such a difference image represents the change in flux distribution, and vortex motion is therefore emphasized. Fig. 1d shows a difference image after a small increase in the applied field. We find in this case that the vortex motion is non-uniform and occurs in isolated jumps as well as collective motion involving many vortices.

With this new improvement, MO-imaging becomes the first technique allowing real-time studies of flux dynamics on scales ranging from individual flux quanta up to global distributions of magnetic induction. Possibilities for future studies include imaging of vortices driven by transport currents, 'shaking' experiments involving AC currents and AC magnetic fields, imaging of vortices interacting with pinning centers, as well as vortex motion across surface barriers. Work is already underway to study the temporal and spatial correlation of vortex motion during flux penetration.

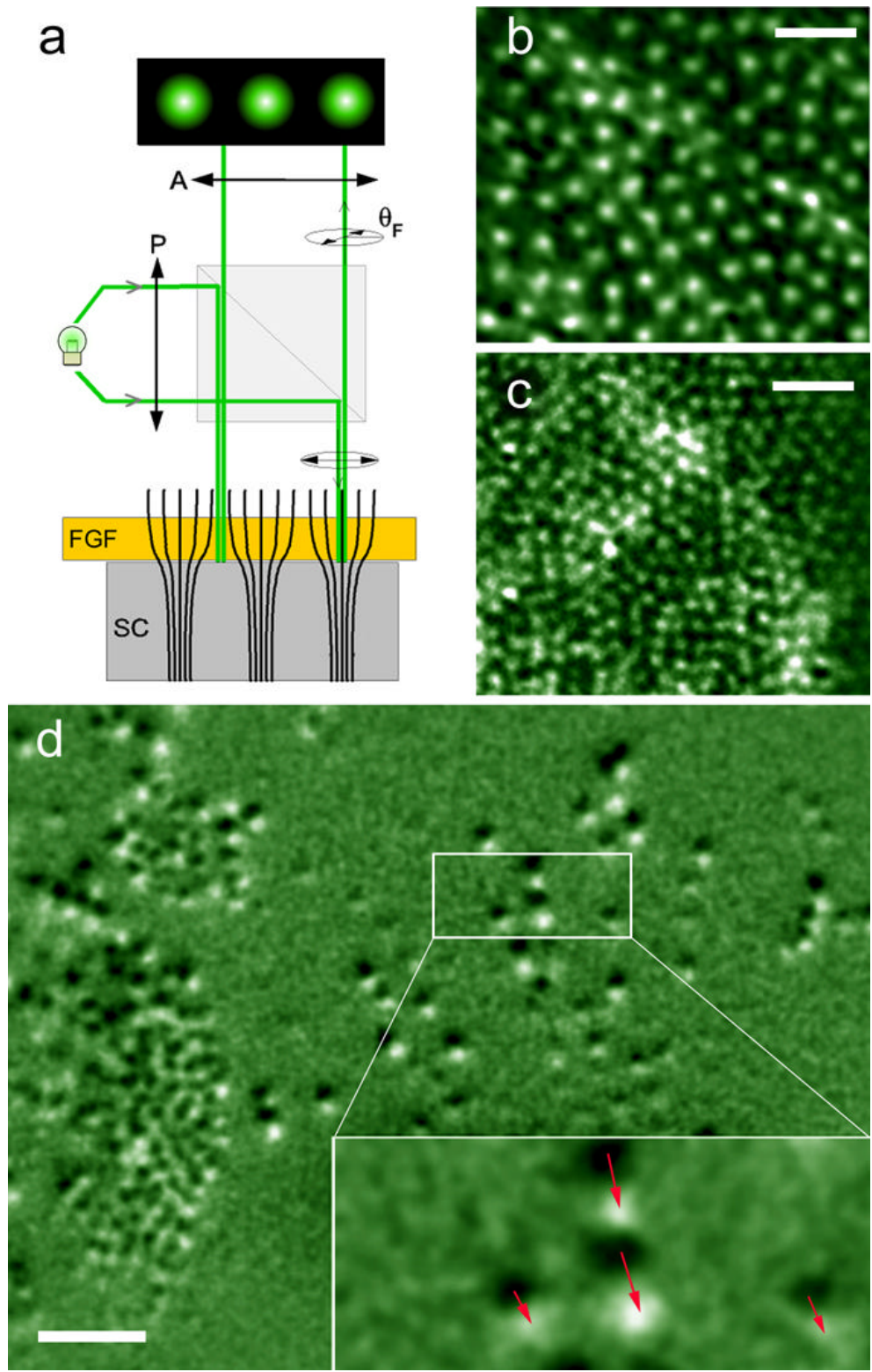

**Figure 1. a**, Principle of MO-imaging. Maxima of the magnetic field from vortices in a superconducting sample (SC) gives maxima in the Faraday rotation $\theta_F$ of incoming plane polarized light in a ferrite garnet layer (FGF) near the sample. Vortices appear as bright spots when imaged using a crossed polarizer(P)/analyser(A) setting. **b** and **c**, MO-images of vortices in a NbSe$_2$ superconducting crystal at 4.0 K after cooling in the earth field ($\approx$1 Oe) and a 3 Oe applied field, respectively. **d**, Vortex dynamics during flux penetration. The image shows the change in flux distribution over a 1 sec. time interval after a 4 mOe increase in the applied field. Dark and bright spots represent initial and final vortex positions, respectively. Medium brightness corresponds to unchanged flux distribution, indicating stationary vortices. The insert shows a close up of four vortex jumps. Arrows indicate the direction of vortex motion. Scalebar 10 μm.